\documentclass[10pt,pra,twocolumn,floatfix]{revtex4}
\usepackage{amssymb}
\usepackage{amsfonts}
\usepackage{amssymb}
\usepackage{amsmath}
\usepackage[dvips]{graphicx}
\usepackage{color}

\begin{document}

\title{Einstein-Podolsky-Rosen steering in Gaussian weighted graph states}
\author{Meihong Wang$^{1,2}$, Xiaowei Deng$^{1,3}$, Zhongzhong Qin$^{1,2}$
and Xiaolong Su$^{1,2}$}
\email{suxl@sxu.edu.cn}
\affiliation{$^{1}$State Key Laboratory of Quantum Optics and Quantum Optics Devices, \\
Institute of Opto-Electronics, Shanxi University, Taiyuan, 030006, People's
Republic of China \\
$^{2}$Collaborative Innovation Center of Extreme Optics, Shanxi University,\\
Taiyuan, Shanxi 030006, People's Republic of China\\
$^{3}$Shenzhen Institute for Quantum Science and Engineering and Department
of Physics, Southern University of Science and Technology, \\
Shenzhen, 518055, People's Republic of China \\
}

\begin{abstract}
Einstein-Podolsky-Rosen (EPR) steering, as one of the most intriguing
phenomenon of quantum mechanics, is a useful quantum resource for quantum
communication. Understanding the type of EPR steering in a graph state is the
basis for application of it in a quantum network. In this paper, we present
EPR steering in a Gaussian weighted graph state, including a linear tripartite
and a four-mode square weighted graph state. The dependence of EPR steering
on weight factor in the weighted graph state is analyzed. Gaussian EPR
steering between two modes of a weighted graph state is presented, which does
not exist in the Gaussian cluster state (where the weight factor is unit).
For the four-mode square Gaussian weighted graph state, EPR steering between
one and its two nearest modes is also presented, which is absent in the
four-mode square Gaussian cluster state. We also show that Gaussian EPR
steering in a weighted graph state is also bounded by the
Coffman-Kundu-Wootters monogamy relation. The presented results are useful
for exploiting EPR steering in a Gaussian weighted graph state as a valuable
resource in multiparty quantum communication tasks.
\end{abstract}

\maketitle

\section{ Introduction}

Einstein-Podolsky-Rosen (EPR) steering, proposed by Schr\"{o}dinger in 1935,
is an intriguing phenomena in quantum mechanics \cite%
{EPR,Schrodinger1,Schrodinger2}. Suppose Alice and Bob share an EPR
entangled state which is separated in space. It allows one party, say
Alice, to steer the state of a distant party, Bob, by exploiting their
shared entanglement \cite{EPR,Schrodinger1,Schrodinger2,Saunders2010}, i.e.,
the state in Bob's station will change instantaneously if Alice makes a
measurement on her state. EPR steering stands between Bell nonlocality \cite%
{Bell} and EPR entanglement \cite{EPREntanglement} and represents a weaker
form of quantum nonlocality in the hierarchy of quantum correlations. EPR
steering can be regarded as verifiable entanglement distribution by an
untrusted party, while entangled states need both parties to trust each
other and Bell nonlocality is valid assuming that they distrust each other 
\cite{WisemanPRA}.

EPR steering has recently attracted increasing interest in quantum optics
and quantum information communities \cite{WisemanPRL,WisemanPRA,Quantifying}%
. Different from entanglement and Bell nonlocality, asymmetric feature is
the unique property of EPR steering \cite%
{WisemanPRL,OneWayNatPhot,OneWayPKLam,OneWayPryde,OneWayGuo}, which is
referred to as one-way EPR steering. In the field of quantum information,
EPR steering has potential applications in one-sided device-independent
quantum key distribution \cite{QKD}, channel discrimination \cite%
{ChannelDiscrimination}, secure quantum teleportation \cite%
{MDReidsecure2013,QHetelep2015}, quantum secret sharing (QSS) \cite%
{Xiang2017}, and remote quantum communication \cite{Wangmh20171,Wangmh20172}.
It has also been shown that the direction of one-way EPR steering can be
actively manipulated \cite{ZhongzhongQin2017}, which may lead to more
consideration in the application of EPR steering. Experimental observation
of multipartite EPR steering has been reported in optical network \cite%
{OneWayPKLam} and photonic qubits \cite{MultipartyCavalcanti,MultipartyJWPan}%
. Very recently, the monogamy relations for EPR steering in a Gaussian cluster
state have been analyzed theoretically in the multipartite state \cite{Xiang2017}
and demonstrated experimentally \cite{Deng2017}.

A graph state is a multipartite entangled state consisting of a set of
vertices connected to each other by edges taking the form of a controlled
phase gate \cite%
{HJBriegel2001,Zhang2006,MHein2004,NCMenicucci2011,JZhang2010}. A cluster
state is a special instance of a graph state where only the neighboring
interaction existed and the weight factor is unit \cite%
{HJBriegel2001,Zhang2006,MHein2004,NCMenicucci2011}. A weighted graph state
describes the state with nonunit weight factor, which denotes the
interaction between vertices \cite{NCMenicucci2011,JZhang2010,PengXue2012}.
The graph state is a basic resource in quantum information and quantum
computation. For example, multiparty Greenberger-Horne-Zeilinger (GHZ) state
and cluster state have been used in quantum communication \cite%
{GHZQSS1,GHZQSS2,GHZQN,GHZTELE,GHZCDC} and one-way quantum computation \cite%
{RRaussendorf2001,Su2013}, respectively.

It has been shown that for some unweighted multipartite entangled state,
Gaussian EPR steering between two modes does not exist, for example, any two
modes in tripartite GHZ state \cite{Reid2013,Coffman2000} and the two
nearest-neighboring modes in four-mode square cluster state \cite{Deng2017}.
It is curious whether EPR steering, which does not exist in an unweighted
state, can be achieved in a weighted graph state. In this paper, we present
the property of EPR steering in a Gaussian weighted graph state, including a
linear tripartite and a four-mode square weighted graph state. By adjusting
the weight factor of the weighted graph state, the dependence of EPR
steering on weight factor is analyzed. EPR steering between two modes, which
is not observed in a tripartite Gaussian GHZ state, is presented in a linear
tripartite weighted graph state. For the four-mode square weighted graph
state, EPR steering between one and its two neighboring modes, which does
not exist in a four-mode square Gaussian cluster state, exists in the
four-mode square weighted graph state. We also show that the CKW-type
monogamy relation is still valid in the Gaussian weighted graph state.
Different from the steerability properties in a previous studied tripartite
and four-mode Gaussian cluster state, which belong to an unweighted graph
state, we observe interesting steerability properties in Gaussian weighted
graph states. These new steerability properties will inspire potential
applications of Gaussian weighted graph states. The existence of EPR
steering in a weighted graph state between any two modes will lead to
a potential security risk when it is applied to implement QSS.

\section{Gaussian EPR steering}

The properties of a ($n_{A}$ and $m_{B}$)-mode Gaussian state of a bipartite
system can be determined by its\textit{\ }covariance matrix 
\begin{equation}
\sigma _{AB}=\left( 
\begin{array}{cc}
A & C \\ 
C^{\top } & B%
\end{array}%
\right) \text{.}
\end{equation}%
with matrix element\textit{\ }$\sigma _{ij}=\langle \hat{\xi}_{i}\hat{\xi}%
_{j}+\hat{\xi}_{j}\hat{\xi}_{i}\rangle /2-\langle \hat{\xi}_{i}\rangle
\langle \hat{\xi}_{j}\rangle $\textit{, }where\textit{\ }$\hat{\xi}\equiv (%
\hat{x}_{1}^{A},\hat{p}_{1}^{A},...,\hat{x}_{n}^{A},\hat{p}_{n}^{A},\hat{x}%
_{1}^{B},\hat{p}_{1}^{B},...,\hat{x}_{m}^{B},\hat{p}_{m}^{B})^{\top }$%
\textit{\ }is the vector of the amplitude and phase quadratures of optical
modes. The submatrixes $A$ and $B$ are corresponding to the reduced states
of Alice's and Bob's subsystems, respectively. The covariance matrix $\sigma
_{AB}$, which corresponds to the optical modes $\hat{A}$ and $\hat{B}$, can
be measured by homodyne detection systems.

The steerability of Bob by Alice ($A\rightarrow B$) for a ($n_{A}+m_{B}$%
)-mode Gaussian state can be quantified by~\cite{Kogias2015} 
\begin{equation}
\mathcal{G}^{A\rightarrow B}(\sigma _{AB})=\max \bigg\{0,\underset{j:\bar{\nu%
}_{j}^{AB\backslash A}<1}{-\sum }\ln (\bar{\nu}_{j}^{AB\backslash A})\bigg\},
\label{eqn:parameter}
\end{equation}%
where $\bar{\nu}_{j}^{AB\backslash A}$ $(j=1,...,m_{B})$ are the symplectic
eigenvalues of $\bar{\sigma}_{AB\backslash A}=B-C^{\top }A^{-1}C$, derived
from the Schur complement of $A$ in\ the covariance matrix $\sigma _{AB}$.
The steerability of Alice by Bob [$\mathcal{G}^{B\rightarrow A}(\sigma
_{AB}) $] can be obtained by swapping the roles of $A$ and $B$.

We analyze tripartite and four-mode steering in a linear tripartite and a
four-mode square Gaussian weighted graph state in the paper. This is done by
using the criterion proposed in Ref. \cite{Kogias2015}, where the
multipartite steering is analyzed by calculating all possible bipartite
separations. In this context, Alice and Bob perform local Gaussian
measurements on their own optical modes.

\section{The Graph state}

A graph state is described by a mathematical graph, that is a set of
vertices connected by edges \cite{MHein2004,NCMenicucci2011,JZhang2010}. A
vertex represents a physical system, e.g., a qubit or a continuous variable
(CV) qumode. An edge between two vertices represents the physical
interaction between the corresponding system. Formally, a weighted graph
state is described by 
\begin{equation}
G=(V, E)
\end{equation}%
of a finite set $V\subset \mathbf{N}$ and a set $E\subset \lbrack V]^{2}$,
the elements of which are subsets of $V$ with two elements each. A finite
set of $n$ vertices $V$ is connected by a set of edges $E$, in which the
strength of interaction is indicated by weight.

Every CV cluster state can be represented by a graph. CV cluster states with
weighted graph are CV stabilizer states, but, different from it, weighted
graph states for qubits are not stabilizer states \cite{NCMenicucci2011}.
Ideal CV cluster states admit a convenient graphical representation in terms
of a symmetric adjacency matrix $\mathbf{C}$ $(\mathbf{C}=\mathbf{C}^{\top 
\text{ }})$, whose $\left( j,k\right) $ entry $C_{jk\text{ }}$is equal to
the weight of the edge linking node $j$ to node $k$ (with no edge
corresponding to a weight of zero) \cite{NCMenicucci2011}. The CV cluster
state associated with graph $\mathbf{C}$ is expressed by \cite%
{NCMenicucci2011} 
\begin{equation}
\left\vert \Psi _{\mathbf{C}}\right\rangle =\exp (\frac{i}{2}\mathbf{\hat{x}}%
^{\top }\mathbf{C\hat{x}})\left\vert 0\right\rangle _{p}^{\otimes N},
\end{equation}%
where $\mathbf{\hat{x}=(\hat{x}_{1},...,\hat{x}_{N})}^{\top }$ is a column
vector of Schr\"{o}dinger-picture position operators. Thus the quadrature
relations (so-called nullifiers) of CV cluster states are expressed by \cite%
{NCMenicucci2011} 
\begin{equation}
\hat{p}_{a}-\sum_{b\in N_{a}}C_{ab}\hat{x}_{b}\rightarrow 0,\forall a\in G
\end{equation}%
where $\hat{x}_{a}=\hat{a}+\hat{a}^{\dag }$ and $\hat{p}_{a}=(\hat{a}-\hat{a}%
^{\dag })/i$ stand for amplitude and phase quadratures of an optical mode $%
\hat{a}$, respectively. The modes of $b\in N_{a}$ are the nearest neighbors
of mode $\hat{a}.$ $C_{ab}$ represents the strength of interaction between
modes $\hat{b}$ and $\hat{a}$. When the $C_{ab}$ is unit, it corresponds to
a standard unweighted cluster state. While the weight factor is not equal to 
$1$,\ it corresponds to a weighted graph state. For an ideal case (infinite
squeezing), the left-hand side trends to zero, so that the state is a
simultaneous zero eigenstate of them (and of any linear combination of them).

\subsection{The linear tripartite weighted graph state}

The graph representation of a linear tripartite weighted graph state is
shown in Fig. 1(a). In the ideal case, quantum correlations of the
tripartite weighted graph state are expressed by 
\begin{align}
& \hat{p}_{A}-C_{AB}\hat{x}_{B}\rightarrow 0,  \notag \\
& \hat{p}_{B}-C_{AB}\hat{x}_{A}-C_{BC}\hat{x}_{C}\rightarrow 0,  \notag \\
& \hat{p}_{C}-C_{BC}\hat{x}_{B}\rightarrow 0,
\end{align}%
where $C_{jk}$ is the weight factor, which represents the strength of
interaction between modes $j$ and $k$. A linear tripartite weighted cluster
state can be prepared by coupling a phase-squeezed and two
amplitude-squeezed states of light on two beam splitters $T_{1}$ and $T_{2}$%
, as shown in Fig. 1(c).

To cancel the effect of antisqueezing noise completely, the weight factors
are required to satisfy the conditions of $C_{AB}=\sqrt{T_{1}}/\sqrt{%
(1-T_{1})(1-T_{2})}$ and $C_{BC}=\sqrt{T_{2}}/\sqrt{1-T_{2}}$, respectively.
Here, the tripartite weighted graph state is prepared by keeping the
transmittance of $T_{1}=1/3$ unchanged and adjusting the transmittance of $%
T_{2}$ as an example. In this case, weight factors are represented by $%
C_{AB}=1/\sqrt{2(1-T_{2})}$ and $C_{BC}=\sqrt{T_{2}}/\sqrt{1-T_{2}}$,
respectively. Thus the quantum correlations between the amplitude and phase
quadratures of the tripartite weighted graph state are expressed by 
\begin{align}
\Delta ^{2}\left( \hat{p}_{A}-C_{AB}\hat{x}_{B}\right) & =\frac{3-2T_{2}}{%
2-2T_{2}}e^{-2r},  \notag \\
\Delta ^{2}\left( \hat{p}_{B}-C_{AB}\hat{x}_{A}-C_{BC}\hat{x}_{C}\right) & =%
\frac{3}{2-2T_{2}}e^{-2r},  \notag \\
\Delta ^{2}\left( \hat{p}_{C}-C_{BC}\hat{x}_{B}\right) & =\frac{1}{1-T_{2}}%
e^{-2r},
\end{align}%
where the subscripts correspond to different optical modes and $\Delta ^{2}$
represents the variance of amplitude or phase quadrature of a quantum state.
When $T_{2}$ is equal to $1/2$, the output state is a tripartite unweighted
cluster state. The details of covariance matrix for the tripartite Gaussian
weighted graph state can be found in Appendix A.

\begin{figure}[tbp]
\begin{center}
\includegraphics[width=80mm]{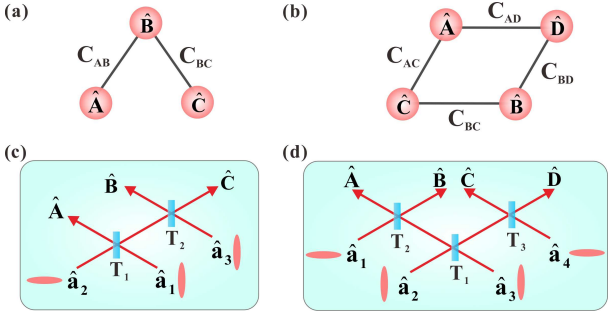}
\end{center}
\caption{Schematic for achieving the weighted graph state. (a) The graph
representation of a linear tripartite Gaussian weighted graph state. (b) The
graph representation of a four-mode square Gaussian weighted graph state.
(c) The schematic for preparing a linear tripartite Gaussian weighted graph
state. (d) The schematic for preparing a four-mode square Gaussian weighted
graph state.}
\label{d}
\end{figure}

\subsection{The four-mode square weighted graph state}

The graph representation of a four-mode square weighted graph state is shown
in Fig. 1(b). In the ideal case, the quadrature correlations of the
four-mode square Gaussian weighted graph state are expressed by 
\begin{align}
& \hat{p}_{A}-C_{AC}\hat{x}_{C}-C_{AD}\hat{x}_{D}\rightarrow 0,  \notag \\
& \hat{p}_{B}-C_{BC}\hat{x}_{C}-C_{BD}\hat{x}_{D}\rightarrow 0,  \notag \\
& \hat{p}_{C}-C_{AC}\hat{x}_{A}-C_{BC}\hat{x}_{B}\rightarrow 0,  \notag \\
& \hat{p}_{D}-C_{AD}\hat{x}_{A}-C_{BD}\hat{x}_{B}\rightarrow 0,
\end{align}%
where $C_{jk\text{ }}$is the strength of interaction between modes $j$ and $%
k $. As shown in Fig. 1(d), the four-mode weighted graph state can be
prepared by coupling two phase-squeezed and two amplitude-squeezed states of
light on an optical beam-splitter network, which consists of three optical
beam splitters with transmittances of $T_{1},T_{2}$ and $T_{3}$. In this
paper, the four-mode weighted graph state is prepared by fixing the
transmittances $T_{1}=1/5$, $T_{3}=1/2$ and adjusting the transmittance of
beam splitter $T_{2}$.

Similarly, to cancel the effect of antisqueezing noise completely, the
weight factors are required to satisfy $C_{AC}=C_{AD}=C_{A}=\sqrt{2T_{2}}$
and $C_{BC}=C_{BD}=C_{B}=\sqrt{2(1-T_{2})}$, respectively. Because the
weight factor between mode $\hat{C}$ and neighboring modes is equal to that
between mode $\hat{D}$ and neighboring modes, modes $\hat{C}$ and $\hat{D}$
are completely symmetric in the four-mode weight graph state.

In this case, the quantum correlations between the amplitude and phase
quadratures of the four-mode square Gaussian weighted graph state are
expressed by 
\begin{eqnarray}
\Delta ^{2}\left( \hat{p}_{A}-C_{A}\hat{x}_{C}-C_{A}\hat{x}_{D}\right)
&=&\left( 1+4T_{2}\right) e^{-2r},  \notag \\
\Delta ^{2}\left( \hat{p}_{B}-C_{B}\hat{x}_{C}-C_{B}\hat{x}_{D}\right)
&=&\left( 5-4T_{2}\right) e^{-2r},  \notag \\
\Delta ^{2}\left( \hat{p}_{C}-C_{A}\hat{x}_{A}-C_{B}\hat{x}_{B}\right)
&=&3e^{-2r},  \notag \\
\Delta ^{2}\left( \hat{p}_{D}-C_{A}\hat{x}_{A}-C_{B}\hat{x}_{B}\right)
&=&3e^{-2r},
\end{eqnarray}%
where the subscripts correspond to different optical modes, $C_{A}$ and $%
C_{B}$ represent the weight factors, i.e., the strength of interaction
between mode $\hat{A}$ and its neighboring mode ($\hat{C}$ or $\hat{D}$),
and that between mode $\hat{B}$ and its neighboring mode ($\hat{C}$ or $\hat{%
D}$), respectively. When $T_{2}=1/2$ is chosen, the state is a four-mode
square Gaussian unweighted graph state. The details of the covariance matrix for
the four-mode square Gaussian weighted graph state can be found in Appendix
B.

\begin{figure}[tbp]
\begin{center}
\includegraphics[width=80mm]{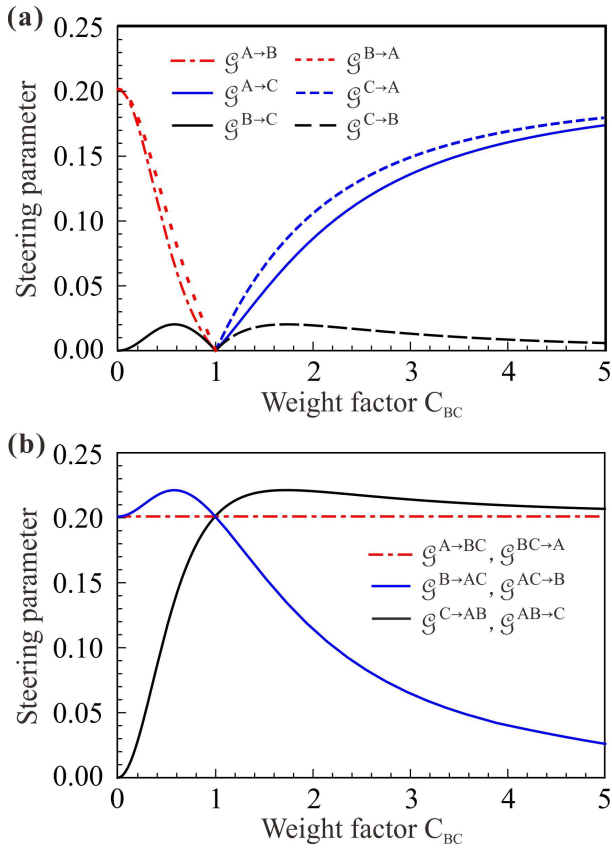}
\end{center}
\caption{Dependence of steering parameter on the weight factor $C_{BC}$ in
the tripartite Gaussian weighted graph state. (a) The pairwise bipartite
steering between any two modes. (b) Steering parameter between one and the
remaining two modes. }
\end{figure}

\section{Results}

\subsection{EPR steering in a linear tripartite weighted graph state}

In the tripartite weighted graph state, EPR steering for (1+1) mode and
(1+2) mode as a function of weight factor $C_{BC}$, as an example, under
Gaussian measurement are shown in Figs. 2(a) and 2(b), respectively, where
the squeezing parameter $r=0.345$ (corresponding to 3 dB squeezing) is
chosen.

As shown in Fig. 2(a), EPR steering between any two modes does not exist in
the condition of $C_{AB}=C_{BC}=1$, which corresponds to a linear tripartite
unweighted graph state. However, EPR steering between any two modes appears
in a Gaussian weighted graph state, which corresponds to the condition of $%
C_{BC}\neq 1$. The steerabilities $\mathcal{G}^{A\rightarrow B},\mathcal{G}%
^{B\rightarrow A}$ and $\mathcal{G}^{B\rightarrow C}$ are larger than zero
in the condition of $C_{BC}<1$ (red lines and black solid line). The other
steerabilities between two modes, including $\mathcal{G}^{A\rightarrow C},%
\mathcal{G}^{C\rightarrow A}$and $\mathcal{G}^{C\rightarrow B}$, exist when
the weight factor is $C_{BC}>1$ (blue and black dashed lines). Especially,
comparing the black solid and dashed lines in Fig. 2(a), we observe one-way
EPR steering between modes $\hat{B}$ and $\hat{C}$ when the weight factor is
fixed. For example, when $C_{BC}=1/2$, only $\mathcal{G}^{B\rightarrow C}$
exist. The steerability $\mathcal{G}^{A\rightarrow C}$ and $\mathcal{G}%
^{A\rightarrow B}$ do not exist in the case of $C_{BC}<1$ and $C_{BC}>1$,
respectively. The reason for absence of the steering is the monogamy
relation obtained from the two-observable EPR criterion \cite{Reid2013}: two
parties cannot steer the third party simultaneously using the same steering
witness. This is the same with the results of the unweighted graph state.
From these results, we clearly see that EPR steering between any two modes,
which does not exist in a linear tripartite CV GHZ state (an unweighted
graph state) \cite{Deng2018}, exists in the tripartite Gaussian weighted
graph state with nonunit weight factor.

The steering parameters between one and the other two modes in the
tripartite Gaussian weighted graph state are shown in Fig. 2(b). The
steerability $\mathcal{G}^{A\rightarrow BC}$\ is not changed, while the
steerabilities $\mathcal{G}^{B\rightarrow AC}$ and $\mathcal{G}%
^{C\rightarrow AB}$ are changed along with the increase of the weight
factor $C_{BC}$. The reason for steerability $\mathcal{G}^{A\rightarrow BC}$%
\ keeping unchanged with the weight factor $C_{BC}$ is that the mode $\hat{A}
$\ is not affected by the transmittance of beam splitter $T_{2}$; only modes 
$\hat{B}$ and $\hat{C}$ are related to the transmittance of beam splitter $%
T_{2}$.

Secret sharing is conventional protocol to distribute a secret message to a
group of parties, who cannot access it individually but have to cooperate in
order to decode it and prevent eavesdropping, for example, if one player
(Bob) can steer the state owned by the dealer (Alice) in a three parties
QSS. Bob may have the ability to decode the secret by himself, and does not
need the collaboration with another player (Claire). In this case, the QSS
will not be secure since one player can obtain the secret independently.

It has been shown that an unweighted tripartite Gaussian cluster state can
be used as the resource of QSS since no steerabilities between any two modes
exist [see the case of $C_{AB}=C_{BC}=1$ in Fig. 1(a)] \cite{Deng2018}.
Here, we have to point out that the potential security risk may exist for
three parties QSS using a linear tripartite Gaussian weighted graph state as
a resource state, due to the existence of EPR steering between two modes. For
example, when the weight factor $C_{BC}=1/2$, the steerabilities $\mathcal{G}%
^{A\rightarrow B},\mathcal{G}^{B\rightarrow A}$ and $\mathcal{G}%
^{B\rightarrow C}$ exist, which means that when any one of modes $\hat{B}$, $%
\hat{A}$ and $\hat{C}$ is chosen as a dealer in QSS, there will be a
security risk that modes $\hat{A}$ and $\hat{B}$ may get the secret alone.
The similar result can be found in the case of $C_{BC}>1$.

\subsection{EPR steering in a four-mode square weighted graph state}

As shown in Fig. 1(d), based on the relation between weight factor and
transmittance, we can achieve a four-mode square weighted graph state by
changing the transmittance $T_{2}$. Because the weight factors have the
relation of $C_{A}^{2}+C_{B}^{2}=2$, the dependence of steering parameters
on the weight factor $C_{A}$ is taken as an example to analyze the steering
parameters of the four-mode weighted graph state. The dependence of EPR
steering on weight factor $C_{A}$ under Gaussian measurements is shown in
Fig. 3,\emph{\ }when the squeezing parameter $r=0.345$ (corresponding to 3
dB squeezing) is chosen.

\begin{figure}[tbp]
\begin{center}
\includegraphics[width=80mm]{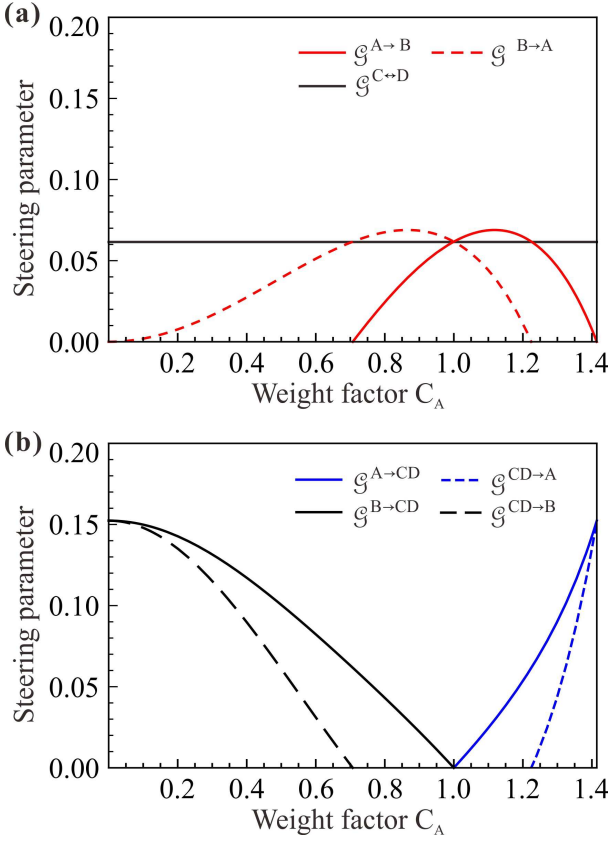}
\end{center}
\caption{Difference of EPR steering between unweighted and weighted
graph state, including (1+1) mode and (1+2) mode, in the four-mode square
Gaussian weighted graph state. (a) Steering parameter between two diagonal
modes. (b) Steering parameter between one mode ($\hat{A}$ or $\hat{B}$) and
a group comprising two nearest-neighboring modes.}
\end{figure}

It has been shown that EPR steering does not exist between any two
neighboring modes and between one mode and the collaboration of its two
neighboring modes in a four-mode square Gaussian unweighted $\left(
C_{jk}=1\right) $ cluster state \cite{Deng2017}. Different from the
unweighted state, one-way EPR steering $\mathcal{G}^{A\rightarrow B}$and $%
\mathcal{G}^{B\rightarrow A}$ is observed in the case of $1.22<C_{A}<1.41$\
and $0<C_{A}<0.71$, respectively [red solid and dashed lines in Fig. 3(a)].
EPR steering between modes $\hat{C}$\ and $\hat{D}$\ is invariable even if
the weight factor $C_{A}$ is changed (black line). This is because the
weighted graph state is obtained by changing the beam splitter $T_{2}$\
between modes $\hat{A}$\ and $\hat{B}$; thus the composition of modes $\hat{C%
}$\ and $\hat{D}$\ is not changed.

We also analyze the steerability between one and its two nearest modes in
the four-mode square Gaussian weighted graph state, which is shown in Fig.
3(b). We can see that the EPR steering between $\hat{A}$ $(\hat{B})$\ and a
group comprising its two nearest-neighboring modes ($\hat{C}$ and $\hat{D}$)
exists in the Gaussian weighted graph state. One-way EPR steering $\mathcal{G}%
^{A\rightarrow CD}$and $\mathcal{G}^{B\rightarrow CD}$\ is observed in the
condition of $1<C_{A}<1.22$\ and $0.71<C_{A}<1$, respectively.

Here, we only present the results that steerabilities of a four-mode square
Gaussian weighted graph state are different from that of a four-mode square
Gaussian unweighted graph state. The details of the steerabilities of a
four-mode square Gaussian unweighted graph state can be found in Ref. \cite%
{Deng2017}. Please note that although the optical mode is not transmitted
over a lossy channel, one-way EPR steering is also presented in the
Gaussian weighted graph state. The reason is that the symmetry is broken in
the Gaussian weighted graph state, just as the previous observed one-way EPR
steering in a lossy channel \cite{Deng2017}.

\subsection{Verification of CKW-type monogamy relation}

\begin{figure}[tbp]
\begin{center}
\includegraphics[width=80mm]{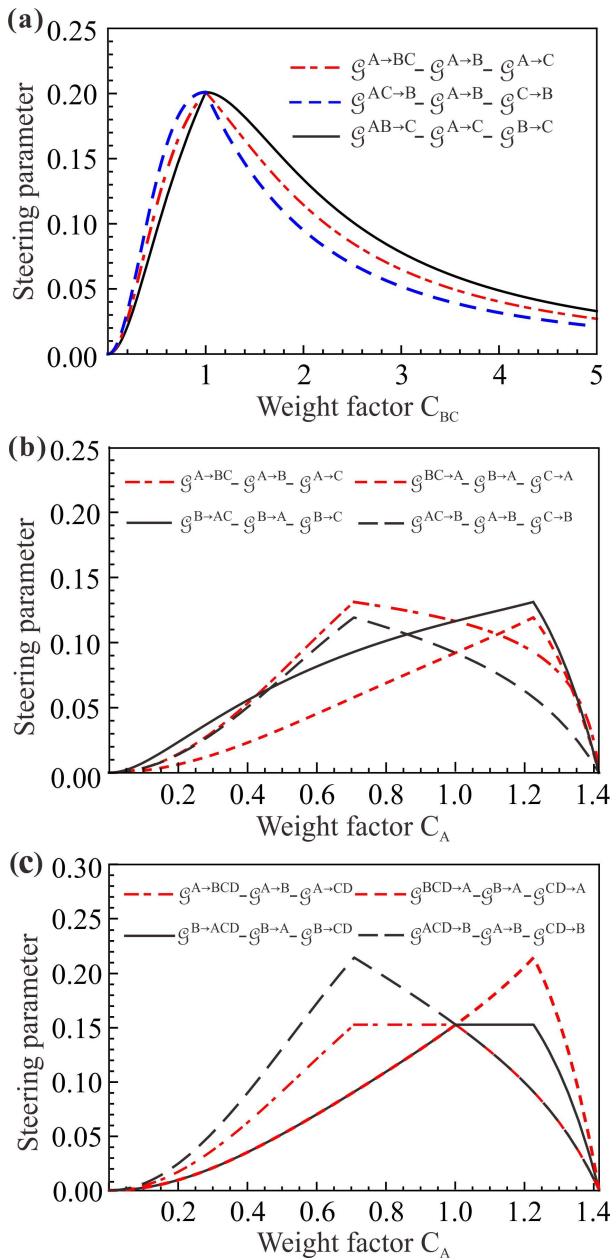}
\end{center}
\caption{Monogamy relation in the Gaussian weighted graph state. (a) The
monogamy relation for all types EPR steering between one and the other two
modes in the tripartite Gaussian weighted graph state. (b),(c) Validation of
generalized CKW-type monogamy for steering in the four-mode square Gaussian
weighted graph state.}
\end{figure}

The Coffman-Kundu-Wootters(CKW)-type monogamy relations \cite{Coffman2000},
which quantify how the steering is distributed among different subsystems 
\cite{Xiang2017}, are expressed by 
\begin{eqnarray}
\mathcal{G}^{k\rightarrow (i,j)}(\sigma _{ijk})-\mathcal{G}^{k\rightarrow
i}(\sigma _{ijk})-\mathcal{G}^{k\rightarrow j}(\sigma _{ijk}) &\geq &0, 
\notag \\
\mathcal{G}^{(i,j)\rightarrow k}(\sigma _{ijk})-\mathcal{G}^{i\rightarrow
k}(\sigma _{ijk})-\mathcal{G}^{j\rightarrow k}(\sigma _{ijk}) &\geq &0,
\end{eqnarray}%
where $i,j,k\in \{\hat{A},\hat{B},\hat{C}\}$ in the tripartite weighted
graph state. Here, we confirm the CKW-type monogamy relation is coincident
for all types of EPR steering in the linear tripartite and four-mode square
Gaussian weighted graph state, as shown in Fig. 4.

Figure. 4(a) shows the CKW-type monogamy relation in the tripartite Gaussian
weighted graph state. When the weight factor $C_{BC}<1$, the CKW-type
monogamy relations $\mathcal{G}^{A\rightarrow BC}-\mathcal{G}^{A\rightarrow
B}-\mathcal{G}^{A\rightarrow C}=$\ $\mathcal{G}^{C\rightarrow AB}-\mathcal{G}%
^{C\rightarrow A}-\mathcal{G}^{C\rightarrow B}$ (red dashed-dotted line), $%
\mathcal{G}^{B\rightarrow AC}-\mathcal{G}^{B\rightarrow A}-\mathcal{G}%
^{B\rightarrow C}=\mathcal{G}^{BC\rightarrow A}-\mathcal{G}^{B\rightarrow A}-%
\mathcal{G}^{C\rightarrow A}=$\ $\mathcal{G}^{AB\rightarrow C}-\mathcal{G}%
^{A\rightarrow C}-\mathcal{G}^{B\rightarrow C}$ (black solid line), and $%
\mathcal{G}^{AC\rightarrow B}-\mathcal{G}^{A\rightarrow B}-\mathcal{G}%
^{C\rightarrow B}$\ (blue dashed line) are valid, respectively. When the
weight factor $C_{BC}>1,$ the CKW-type monogamy relations $\mathcal{G}%
^{A\rightarrow BC}-\mathcal{G}^{A\rightarrow B}-\mathcal{G}^{A\rightarrow C}=
$\ $\mathcal{G}^{B\rightarrow AC}-\mathcal{G}^{B\rightarrow A}-\mathcal{G}%
^{B\rightarrow C}$ (red dashed-dotted line), $\mathcal{G}^{C\rightarrow AB}-%
\mathcal{G}^{C\rightarrow A}-\mathcal{G}^{C\rightarrow B}=\mathcal{G}%
^{BC\rightarrow A}-\mathcal{G}^{B\rightarrow A}-\mathcal{G}^{C\rightarrow A}=
$\ $\mathcal{G}^{AC\rightarrow B}-\mathcal{G}^{A\rightarrow B}-\mathcal{G}%
^{C\rightarrow B}$ (blue dashed line), and $\mathcal{G}^{AB\rightarrow C}-%
\mathcal{G}^{A\rightarrow C}-\mathcal{G}^{B\rightarrow C}$\ (black solid
line) are also valid, respectively.

We also confirm the general monogamy relations in the four-mode square
Gaussian weighted graph state \cite{genemono}, especially the steerabilities
that are different from that of the unweighted graph state, are valid, as
shown in Figs. 4(b)-4(c), where $i,j,k\in \{\hat{A},\hat{B},\hat{C},\hat{D}%
\}$ or $\{\hat{A},\hat{C}\hat{D},\hat{B}\}$. The CKW-type monogamy
relations, including EPR steering between modes $\hat{A}\ $and $\hat{B}$,
are shown in Fig. 4(b). Due to the symmetry of modes $\hat{C}\ $ and $\hat{D}
$, the validation of monogamy relations for mode $\hat{C}\ $ are suitable
for $\hat{D}$.\ The generalized CKW-type monogamy relations are also valid,
as shown in Fig. 4(c).

When $i,j,k\in \{\hat{A},\hat{C},\hat{D}\}$ are chosen, the steerabilities $%
\mathcal{G}^{A\rightarrow CD}$ and $\mathcal{G}^{CD\rightarrow A}$ exist for
the four-mode square Gaussian weighted graph state as shown in Fig. 3(b).
The EPR steering between modes $\hat{A}\ $ and $\hat{C}(\hat{D})$\ for
the four-mode square weighted graph state does not exist, i.e., $\mathcal{G}%
^{A\rightarrow C}=0$ and $\mathcal{G}^{C\rightarrow A}=0$, which is the same
as that of the four-mode square unweighted graph state as shown in Ref. \cite%
{Deng2017}. The CKW-type monogamy relations $\mathcal{G}^{A\rightarrow CD}-%
\mathcal{G}^{A\rightarrow C}-\mathcal{G}^{A\rightarrow D}\geq 0$ and $%
\mathcal{G}^{CD\rightarrow A}-\mathcal{G}^{C\rightarrow A}-\mathcal{G}%
^{D\rightarrow A}\geq 0$ are always valid. The same results are obtained for
steerabilities among mode $\hat{B}$ and modes ($\hat{C}$, $\hat{D}$).

\section{Discussion and conclusion}

EPR steering analyzed in this paper are arbitrary bipartite separations of a
tripartite and four-mode Gaussian weighted graph states based on the
necessary and sufficient criterion under Gaussian measurements \cite%
{Kogias2015}, which
quantifies EPR steering for bipartite separations of a multipartite Gaussian
state. A quantum system involving more than three subsystems has different
possible partitions, and it has been discussed for entanglement \cite%
{Multient1,Multient2,Multient3} and Bell nonlocality \cite%
{MultiBell1,MultiBell2}. For EPR steering of Gaussian states, the criterion
which is used to quantify the quantum steering for multipartition of a
multipartite Gaussian state remains an open question until now and it is
worthy of further investigation.

The study on quantum nonlocality and EPR steering has deepened our
understanding of the foundation of quantum theory. Recently, postquantum
nonlocality has been discussed in discrete \cite{Post1} and continuous
variable scenarios \cite{Post2,Post3}. The postquantum steering has been
studied in a discrete scenario \cite{Post4,Post5}. However, the postquantum
steering for a continuous variable system has not been discussed, which
remains an open question.

In this paper, the quantum states are Gaussian states of a continuous variable
system and the measurements are Gaussian measurements. The necessary and
sufficient criterion for EPR steering of a Gaussian state proposed in Ref.
\cite{Kogias2015} is used to quantify the EPR steering in Gaussian weighted graph states.Recently, it has been shown that non-Gaussian measurements can lead to extra
steerability even for Gaussian states \cite{OneWayPryde,nonGaussian2}, and
might allow for circumventing some monogamy constraints \cite%
{Reid2013,nonGaussian3,nonGaussian4}. It will be interesting to investigate
EPR steering in Gaussian weighted graph states with non-Gaussian
measurements.

In conclusion, steering parameters in a linear tripartite and a four-mode
square Gaussian weighted graph state are presented. Comparing with the
unweighted graph state, we conclude that a weighted graph state features
richer steering properties. EPR steering that is absent in the Gaussian
unweighted graph state is presented in the Gaussian weighted graph state.
The pairwise bipartite steering exists in the tripartite Gaussian weighted
graph state. EPR steering between one and its two nearest modes is also
observed in the four-mode square Gaussian weighted graph state, which could
not be obtained in the four-mode square Gaussian unweighted graph state. We
also show that the CKW-type monogamy relations are valid in the Gaussian
weighted graph states.

We also analyze quantum entanglement in the linear tripartite and four-mode
square Gaussian weighted graph state. Different from the quantum steering,
quantum entanglement is always maintained in the linear tripartite and
four-mode square Gaussian weighted graph state. This result is the same as that
obtained in Ref. \cite{JZhang2010}, where the entanglement of the weighted
graph state is analyzed.

QSS can be implemented when the players are separated in a local
quantum network and collaborate to decode the secret sent by the dealer who
owns the other one mode \cite{GHZQSS1}. In this case, the dealer must not be
steered by any one of two players; only the collective steerability is
needed. Thus the presence of EPR steering between any two modes in a
linear tripartite Gaussian weighted graph state shows that the Gaussian
weighted graph state is not a good resource for QSS.

There are other suitable quantum information tasks using EPR steering in a
Gaussian weighted graph state as a resource. For example, for the tripartite
Gaussian weighted graph state with weight factor $0<C_{BC}<1$, the
steerability between modes $\hat{A}$ and $\hat{B}$ always exists, and only
the steerability from $\hat{B}$ to $\hat{C}$ exists. In this case, the state
can be used as resource state of quantum conference \cite%
{Conference1,Conference2}. Especially, the user B (who owns mode $\hat{B}$)
can send information to users A (who owns mode $\hat{A}$) and C (who owns
mode $\hat{C}$), and the communication between users B and C is one-way
since only steerability from modes $\hat{B}$ to $\hat{C}$ exists. Thus this
kind of quantum conference based on the tripartite Gaussian weighted graph
state is one-way quantum conference, in which only user B can send
information to users A and C, while users A and C can not send information
to user B. \bigskip

\section*{ACKNOWLEDGMENTS}

This research was supported by the NSFC (Grants No. 11834010, No.11904160, and No.
61601270), the program of Youth Sanjin Scholar, National Key R\&D Program of China (Grant No. 2016YFA0301402), and the Fund for
Shanxi "1331 Project" Key Subjects Construction.

\section*{APPENDIX}

\subsection{Preparation scheme of the linear tripartite Gaussian weighted
graph state}

In this appendix, we present details of the preparation scheme for the
tripartite Gaussian weighted graph state. As shown in Fig. 1(c) in the main
text, the tripartite Gaussian weighted graph state is prepared by coupling
three squeezed states on two optical beam splitters $T_{1}$ and $T_{2}$.
Three input squeezed states are expressed by 
\begin{align}
\hat{a}_{1}& =e^{-r_{1}}\hat{x}_{1}^{(0)}+ie^{r_{1}}\hat{p}_{1}^{(0)}, 
\notag \\
\hat{a}_{2}& =e^{r_{2}}\hat{x}_{2}^{(0)}+ie^{-r_{2}}\hat{p}_{2}^{(0)}, 
\notag \\
\hat{a}_{3}& =e^{-r_{3}}\hat{x}_{3}^{(0)}+ie^{r_{3}}\hat{p}_{3}^{\left(
0\right) },  \tag{A1}
\end{align}%
where $r_{i}$ ($i=1,2,3$) is the squeezing parameter, and the superscript of
the amplitude and phase quadratures represent the vacuum state. Under this
notation, the variances of amplitude and phase quadratures for vacuum state
are $\Delta ^{2}\hat{x}^{(0)}=\Delta ^{2}\hat{p}^{(0)}=1$. The
transformation matrix of the beam-splitter network is 
\begin{equation}
U_{1}=\left[ 
\begin{array}{ccc}
-\sqrt{T_{1}} & -\sqrt{1-T_{1}} & 0 \\ 
i\sqrt{(1-T_{1})(1-T_{2})} & -i\sqrt{T_{1}(1-T_{2})} & \sqrt{T_{2}} \\ 
-\sqrt{(1-T_{1})T_{2}} & \sqrt{T_{1}T_{2}} & -i\sqrt{1-T_{2}}%
\end{array}%
\right] ,  \tag{A2}
\end{equation}

After the conversion of the beam-splitter network, the output modes are
given by 
\begin{align}
\hat{A}& =-\sqrt{T_{1}}\hat{a}_{1}-\sqrt{1-T_{1}}\hat{a}_{2},  \notag \\
\hat{B}& =i\sqrt{(1-T_{1})(1-T_{2})}\hat{a}_{1}-i\sqrt{T_{1}(1-T_{2})}\hat{a}%
_{2}+\sqrt{T_{2}}\hat{a}_{3},  \notag \\
\hat{C}& =-\sqrt{(1-T_{1})T_{2}}\hat{a}_{1}+\sqrt{T_{1}T_{2}}\hat{a}_{2}-i%
\sqrt{1-T_{2}}\hat{a}_{3},  \tag{A3}
\end{align}%
respectively. Here, we assume that the squeezed parameters of all the
squeezed states are equal $(r_{1}=r_{2}=r_{3}=r)$.

The Gaussian state can be completely characterized by a covariance matrix. Based
on the expressions of input and output states, the covariance matrix of the
tripartite Gaussian weighted graph state is expressed by 
\begin{equation}
\sigma _{ABC}=\left[ 
\begin{array}{ccc}
\sigma _{A} & f\mathbf{\Omega } & g\mathbf{Z} \\ 
f\mathbf{\Omega } & \sigma _{B} & h\mathbf{\Omega } \\ 
g\mathbf{Z} & h\mathbf{\Omega } & \sigma _{C}%
\end{array}%
\right] ,  \tag{A4}
\end{equation}%
where$\ \ \ \ \ \ \ \ $

$\ \ \ \ \ \ \ \ \ \ \ \ \ \ \ f=\frac{\sqrt{2(1-T_{2})}(e^{2r}-e^{-2r})}{3}%
, $

$\ \ \ \ \ \ \ \ \ \ \ \ \ \ \ g=\frac{\sqrt{2T_{2}}(e^{-2r}-e^{2r})}{3},\ \
\ \ \ \ \ \ \ \ \ \ \ \ \ \ $

$\ \ \ \ \ \ \ \ \ \ \ \ \ \ \ h=\frac{2\sqrt{T_{2}(1-T_{2})}(e^{2r}-e^{-2r})%
}{3},$

$\ \ \ \ \ \ \mathbf{\Omega =}%
\begin{pmatrix}
0 & 1 \\ 
1 & 0%
\end{pmatrix}%
,\qquad \mathbf{Z}=%
\begin{pmatrix}
1 & 0 \\ 
0 & -1%
\end{pmatrix}%
,$

$\ \ \ \sigma _{A}=\left( 
\begin{array}{cc}
\frac{1}{3}e^{-2r}+\frac{2}{3}e^{2r} & 0 \\ 
0 & \frac{2}{3}e^{-2r}+\frac{1}{3}e^{2r}%
\end{array}%
\right) ,$

$\sigma _{B}=\left( 
\begin{array}{cc}
\frac{2(1-T_{2})e^{2r}+(1+2T_{2})e^{-2r}}{3} & 0 \\ 
0 & \frac{2(1-T_{2})e^{-2r}+(1+2T_{2})e^{2r}}{3}%
\end{array}%
\right) ,$\bigskip

$\sigma _{C}=\left( 
\begin{array}{cc}
\frac{3-2T_{2}}{3}e^{2r}+\frac{2T_{2}}{3}e^{-2r} & 0 \\ 
0 & \frac{3-2T_{2}}{3}e^{-2r}+\frac{2T_{2}}{3}e^{2r}%
\end{array}%
\right) ,$ respectively.$\ \ \ \ \ $

\subsection{Preparation scheme of the four-mode square Gaussian weighted
graph state}

As shown in Fig. 1(d) in the main text, the four-mode square Gaussian
weighted graph state is prepared by coupling four squeezed states on an
optical beam-splitter network. Four input squeezed states are expressed by 
\begin{align}
\hat{a}_{1}& =e^{r_{1}}\hat{x}_{1}^{(0)}+ie^{-r_{1}}\hat{p}_{1}^{(0)}, 
\notag \\
\hat{a}_{2}& =e^{-r_{2}}\hat{x}_{2}^{(0)}+ie^{r_{2}}\hat{p}_{2}^{(0)}, 
\notag \\
\hat{a}_{3}& =e^{-r_{3}}\hat{x}_{3}^{(0)}+ie^{r_{3}}\hat{p}_{3}^{\left(
0\right) },  \notag \\
\hat{a}_{4}& =e^{r_{4}}\hat{x}_{4}^{(0)}+ie^{-r_{4}}\hat{p}_{4}^{\left(
0\right) },  \tag{A5}
\end{align}%
When the transmittances of $T_{1}=1/5$ and $T_{3}=1/2$ are chosen, the
transformation matrix of the beam-splitter network is given by 
\begin{equation}
U_{2}=\left[ 
\begin{array}{cccc}
-\sqrt{1-T_{2}} & -2\sqrt{\frac{T_{2}}{5}} & -i\sqrt{\frac{T_{2}}{5}} & 0 \\ 
\sqrt{T_{2}} & -2\sqrt{\frac{1-T_{2}}{5}} & -i\sqrt{\frac{1-T_{2}}{5}} & 0
\\ 
0 & \frac{i}{\sqrt{10}} & \sqrt{\frac{2}{5}} & -\frac{1}{\sqrt{2}} \\ 
0 & \frac{i}{\sqrt{10}} & \sqrt{\frac{2}{5}} & \frac{1}{\sqrt{2}}%
\end{array}%
\right] ,  \tag{A6}
\end{equation}

Thus the output modes from the optical beam-splitter network are expressed
by 
\begin{align}
\hat{A}& =-\sqrt{1-T_{2}}\hat{a}_{1}-2\sqrt{\frac{T_{2}}{5}}\hat{a}_{2}-i%
\sqrt{\frac{T_{2}}{5}}\hat{a}_{3},  \notag \\
\hat{B}& =\sqrt{T_{2}}\hat{a}_{1}-2\sqrt{\frac{1-T_{2}}{5}}\hat{a}_{2}-i%
\sqrt{\frac{1-T_{2}}{5}}\hat{a}_{3},  \notag \\
\hat{C}& =\frac{i}{\sqrt{10}}\hat{a}_{2}+\sqrt{\frac{2}{5}}\hat{a}_{3}-\frac{%
1}{\sqrt{2}}\hat{a}_{4},  \notag \\
\hat{D}& =\frac{i}{\sqrt{10}}\hat{a}_{2}+\sqrt{\frac{2}{5}}\hat{a}_{3}+\frac{%
1}{\sqrt{2}}\hat{a}_{4},  \tag{A7}
\end{align}%
respectively. Here, we assume that the squeezed parameters of all the
squeezed states are equal $(r_{1}=r_{2}=r_{3}=r_{4}=r)$.

According to the information of input and output states, the covariance
matrix of the four-mode square Gaussian weighted graph state is expressed by 
\begin{equation}
\sigma _{ABCD}=\left[ 
\begin{array}{cccc}
\sigma _{A} & l\mathbf{Z} & m\mathbf{\Omega } & s\mathbf{\Omega } \\ 
l\mathbf{Z} & \sigma _{B} & n\mathbf{\Omega } & v\mathbf{\Omega } \\ 
m\mathbf{\Omega } & n\mathbf{\Omega } & \sigma _{C} & w\mathbf{Z} \\ 
s\mathbf{\Omega } & v\mathbf{\Omega } & w\mathbf{Z} & \sigma _{D}%
\end{array}%
\right] .  \tag{A8}
\end{equation}%
where \ 

$\ \ \ \ \ \ \ \ \ \ \ \ l=\frac{4\sqrt{T_{2}(1-T_{2})}(e^{-2r}-e^{2r})}{5},$

$\ \ \ \ \ \ \ \ \ \ \ m=\frac{\sqrt{2T_{2}}(e^{2r}-e^{-2r})}{5},$

$\ \ \ \ \ \ \ \ \ \ \ \ n=\frac{\sqrt{2(1-T_{2})}(e^{2r}-e^{-2r})}{5},$

$\ \ \ \ \ \ \ \ \ \ \ \ s=\frac{\sqrt{2T_{2}}(e^{2r}-e^{-2r})}{5},$

\ $\ \ \ \ \ \ \ \ \ \ \ v=\frac{\sqrt{2(1-T_{2})}(e^{2r}-e^{-2r})}{5},$

$\ \ \ \ \ \ \ \ \ \ \ \ w=\frac{2(e^{-2r}-e^{2r})}{5},$

$\sigma _{A}=\left( 
\begin{array}{cc}
\frac{5-4T_{2}}{5}e^{2r}+\frac{4T_{2}}{5}e^{-2r} & 0 \\ 
0 & \frac{5-4T_{2}}{5}e^{-2r}+\frac{4T_{2}}{5}e^{2r}%
\end{array}%
\right) ,$

$\sigma _{B}=\left( 
\begin{array}{cc}
\frac{(1+4T_{2})e^{2r}+4(1-T_{2})e^{-2r}}{5} & 0 \\ 
0 & \frac{(1+4T_{2})e^{-2r}+4(1-T_{2})e^{2r}}{5}%
\end{array}%
\right) ,$

$\sigma _{C}=\left( 
\begin{array}{cc}
\frac{3}{5}e^{2r}+\frac{2}{5}e^{-2r} & 0 \\ 
0 & \frac{3}{5}e^{-2r}+\frac{2}{5}e^{2r}%
\end{array}%
\right) ,$

$\sigma _{D}=\left( 
\begin{array}{cc}
\frac{3}{5}e^{2r}+\frac{2}{5}e^{-2r} & 0 \\ 
0 & \frac{3}{5}e^{-2r}+\frac{2}{5}e^{2r}%
\end{array}%
\right) ,$

respectively.

Based on the covariance matrices of the linear tripartite and the four-mode
square Gaussian weighted graph states, the property of the weighted graph
states can be verified.


\begin{thebibliography}{99}
\bibitem{EPR} A. Einstein, B. Podolsky, and N. Rosen, Phys. Rev. \textbf{47}%
, 777 (1935).

\bibitem{Schrodinger1} E. Schr\"{o}dinger, Proc. Cambridge Philos. Soc. 
\textbf{31}, 555 (1935).

\bibitem{Schrodinger2} E. Schr\"{o}dinger, Proc. Cambridge Philos. Soc. 
\textbf{32}, 446 (1936).

\bibitem{Saunders2010} D. J. Saunders, S. J. Jones, H. M. Wiseman, and G. J.
Pryde, Nat. Phys. \textbf{6}, 845 (2010).

\bibitem{Bell} J. S. Bell, Physics \textbf{1}, 195 (1964).

\bibitem{EPREntanglement} R. Horodecki, P. Horodecki, M. Horodecki, and K.
Horodecki, Rev. Mod. Phys. \textbf{81}, 865 (2009).

\bibitem{WisemanPRA} S. J. Jones, H. M. Wiseman, and A. C. Doherty, Phys.
Rev. A \textbf{76}, 052116 (2007).

\bibitem{Quantifying} P. Skrzypczyk, M. Navascu\'{e}s, and D. Cavalcanti,
Phys. Rev. Lett. \textbf{112}, 180404 (2014).

\bibitem{WisemanPRL} H. M. Wiseman, S. J. Jones, and A. C. Doherty, Phys.
Rev. Lett. \textbf{98}, 140402 (2007).

\bibitem{OneWayNatPhot} V. H\"{a}ndchen, T. Eberle, S. Steinlechner, A.
Samblowski, T. Franz, R. F. Werner, and R. Schnabel, Nat. Photon. \textbf{6%
}, 596 (2012).

\bibitem{OneWayPKLam} S. Armstrong, M. Wang, R. Y. Teh, Q. Gong, Q. He, J.
Janousek, H. A. Bachor, M. D. Reid, and P. K. Lam, Nat. Phys. \textbf{11},
167 (2015).

\bibitem{OneWayPryde} S. Wollmann, N. Walk, A. J. Bennet, H. M. Wiseman, and
G. J. Pryde, Phys. Rev. Lett. \textbf{116}, 160403 (2016).

\bibitem{OneWayGuo} K. Sun, X.-J. Ye, J.-S. Xu, X.-Y. Xu, J.-S. Tang, Y.-C. Wu, J.-L. Chen, 
C.-F.Li, and G.-C. Guo, Phys. Rev. Lett. \textbf{116}, 160404 (2016).

\bibitem{QKD} C. Branciard, E. G. Cavalcanti, S. P. Walborn, V. Scarani, and
H. M. Wiseman, Phys. Rev. A \textbf{85}, 010301 (2012).

\bibitem{ChannelDiscrimination} M. Piani and J. Watrous, Phys. Rev. Lett. 
\textbf{114}, 060404 (2015).

\bibitem{MDReidsecure2013} M. D. Reid, Phys. Rev. A \textbf{88}, 062338
(2013).

\bibitem{QHetelep2015} Q. He, L. Rosales-Z\'{a}rate, G. Adesso, and M. D.
Reid, Phys. Rev. Lett. \textbf{115}, 180502 (2015).

\bibitem{Xiang2017} Y. Xiang, I. Kogias, G. Adesso, and Q. He, Phys. Rev. A 
\textbf{95}, 010101(R) (2017).

\bibitem{Wangmh20171} M. Wang, Z. Qin, and X. Su, Phys. Rev. A \textbf{95},
052311 (2017).

\bibitem{Wangmh20172} M. Wang, Z. Qin, Y. Wang, and X. Su, Phys. Rev. A 
\textbf{96}, 022307 (2017).

\bibitem{ZhongzhongQin2017} Z. Qin, X. Deng, C. Tian, M. Wang, X. Su, C.
Xie, and K. Peng, Phys. Rev. A \textbf{95}, 052114 (2017).

\bibitem{MultipartyCavalcanti} D. Cavalcanti, P. Skrzypczyk, G. H. Aguilar,
R. V. Nery, P. H. Souto Ribeiro, and S. P. Walborn, Nat. Commun. \textbf{6},
7941 (2015).

\bibitem{MultipartyJWPan} C. M. Li, K. Chen, Y. N. Chen, Q. Zhang, Y. A.
Chen, and J. W. Pan, Phys. Rev. Lett. \textbf{115}, 010402 (2015).

\bibitem{Deng2017} X. Deng, Y. Xiang, C. Tian, G. Adesso, Q. He, Q. Gong, X.
Su, C. Xie, and K. Peng, Phys. Rev. Lett. \textbf{118}, 230501 (2017).

\bibitem{HJBriegel2001} H. J. Briegel and R. Raussendorf, Phys. Rev. Lett. 
\textbf{86}, 910 (2001).

\bibitem{Zhang2006} J. Zhang and S. L. Braunstein, Phys. Rev. A \textbf{73},
032318 (2006).

\bibitem{MHein2004} M. Hein, J. Eisert, and H. J. Briegel, Phys. Rev. A 
\textbf{69}, 062311 (2004).

\bibitem{NCMenicucci2011} N. C. Menicucci, S. T. Flammia, and P. van Loock,
Phys. Rev. A \textbf{83}, 042335 (2011).

\bibitem{JZhang2010} J. Zhang, Phys. Rev. A \textbf{82}, 034303 (2010).

\bibitem{PengXue2012} P. Xue, Phys. Rev. A \textbf{86}, 023812 (2012).

\bibitem{GHZQSS1} I. Kogias, Y. Xiang, Q. He, and G. Adesso, Phys. Rev. A 
\textbf{95}, 012315 (2017).

\bibitem{GHZQSS2} M. Wang, Y. Xiang, Q. He, and Q. Gong, Phys. Rev. A 
\textbf{91}, 012112 (2015).

\bibitem{GHZQN} P. van Loock and S. L. Braunstein, Phys. Rev. Lett. \textbf{%
84}, 3482 (2000).

\bibitem{GHZTELE} H. Yonezawa, T. Aoki, and A. Furusawa, Nature \textbf{431}%
, 430 (2004).

\bibitem{GHZCDC} J. Jing, J. Zhang, Y. Yan, F. Zhao, C. Xie, and K. Peng,
Phys. Rev. Lett. \textbf{90}, 167903 (2003).

\bibitem{RRaussendorf2001} R. Raussendorf and H. J. Briegel, Phys. Rev.
Lett. \textbf{86}, 5188 (2001).

\bibitem{Su2013} X. Su, S. Hao, X. Deng, L. Ma, M. Wang, X. Jia, C. Xie, and
K. Peng, Nat. Commun. \textbf{4}, 2828 (2013).

\bibitem{Reid2013} M. D. Reid, Phys. Rev. A \textbf{88}, 062108 (2013).

\bibitem{Coffman2000} V. Coffman, J. Kundu, and W. K. Wootters, Phys. Rev. A 
\textbf{61}, 052306 (2000).



\bibitem{Kogias2015} I. Kogias, A. R. Lee, S. Ragy, and G. Adesso, Phys.
Rev. Lett. \textbf{114}, 060403 (2015).

\bibitem{Deng2018} X. Deng, C. Tian, M. Wang, Z. Qin, and X. Su, Opt.
Commun. \textbf{421}, 14 (2018).

\bibitem{genemono} L. Lami, C. Hirche, G. Adesso, and A. Winter, Phys. Rev.
Lett. \textbf{117}, 220502 (2016).

\bibitem{Multient1} J. Sperling and W. Vogel, Phys. Rev. Lett. \textbf{111},
110503 (2013).

\bibitem{Multient2} S. Gerke, J. Sperling, W. Vogel, Y. Cai, J. Roslund, N.
Treps, and C. Fabre, Phys. Rev. Lett. \textbf{114}, 050501 (2015).

\bibitem{Multient3} Z. Qin, M. Gessner, Z. Ren, X. Deng, D. Han, W. Li, X.
Su, A. Smerzi and K. Peng, npj Quantum Inf. \textbf{5}, 3 (2019).

\bibitem{MultiBell1} S. Sami and I. Chakrabarty, and A. Chaturvedi, Phys.
Rev. A \textbf{96}, 022121 (2017).

\bibitem{MultiBell2} M. C. Tran, R. Ramanathan, M. McKague, D. Kaszlikowski,
and T. Paterek, Phys. Rev. A \textbf{98}, 052325 (2018).

\bibitem{Post1} S. Popescu t and D. Rohrlich, Found.Phys. \textbf{%
24}, 379 (1994).

\bibitem{Post2} A. Ketterer, A. Laversanne-Finot, and L. Aolita, Phys. Rev.
A \textbf{97}, 012133 (2018).

\bibitem{Post3} P. Cherian J, A. Mukherjee, A. Roy, S. S. Bhattacharya,
and M. Banik, Phys. Rev. A \textbf{99}, 012105 (2019).

\bibitem{Post4} A. B. Sainz, N. Brunner, D. Cavalcanti, P.
Skrzypczyk, and T. V\'{e}rtesi, Phys. Rev. Lett. \textbf{115}, 190403 (2015).

\bibitem{Post5} M. Banik, J. Math. Phys. \textbf{56}, 052101 (2015).

\bibitem{nonGaussian2} S. -W. Ji, J. Lee, J. Park, and H. Nha, Sci. Rep. 
\textbf{6}, 29729 (2016).

\bibitem{nonGaussian3} S. -W. Ji, M. S. Kim, and H. Nha, J. Phys. A: Math.
Theor. \textbf{48}, 135301 (2015).

\bibitem{nonGaussian4} G. Adesso and R. Simon, J. Phys. A: Math. Theor. 
\textbf{49}, 34LT02 (2016).

\bibitem{Conference1} Y. Wu, J. Zhou, X. Gong, Y. Guo, Z.-M. Zhang, and G. He,
Phys. Rev. A \textbf{93}, 022325 (2016).

\bibitem{Conference2} Y. Wang, C. Tian, Q. Su, M. Wang, X. Su, Sci. Chin.
Inf. Sci. \textbf{62}, 072501 (2019).
\end{thebibliography}
\end{document}